\documentclass[a4paper,10pt,twocolumn]{revtex4}
\usepackage{amssymb}
\usepackage[pdftex]{graphicx}	
	\DeclareGraphicsExtensions{.pdf,.png,.jpg}
\usepackage[applemac]{inputenc}
\usepackage{amsmath,amssymb,amsthm}
\usepackage{mathbbol,bbm}
\usepackage{graphicx,float}
\usepackage[final]{showkeys}
\usepackage{bm,dsfont}
\usepackage[abs]{overpic}
\usepackage{booktabs}
\usepackage{color}
\usepackage{braket}

\newcommand{\beq}{\begin{eqnarray}}
\newcommand{\eeq}{\end{eqnarray}}

\def\ket#1{|#1\rangle}

\begin{document}
\title{Scalable Fermionic Error Correction in Majorana Surface Codes}
\author{Oscar Viyuela$^{1,2}$, Sagar Vijay$^{1,2}$, Liang Fu$^{1}$}
\affiliation{1. Department of Physics, Massachusetts Institute of Technology, Cambridge, MA 02139, USA\\
2. Department of Physics, Harvard University, Cambridge, MA 02318, USA }

\vspace{-3.5cm}

\begin{abstract}
We study the error correcting properties of Majorana Surface Codes (MSC), topological quantum codes constructed out of interacting Majorana fermions, which can be used to store quantum information and perform quantum computation. These quantum memories suffer from purely ``fermionic" errors, such as quasiparticle poisoning (QP), that have no analog in conventional platforms with bosonic qubits. In physical realizations where QP dominates, we show that errors can be corrected provided that the poisoning rate is below a threshold of $\sim11\%$. When QP is highly suppressed and fermionic bilinear (``bosonic") errors become dominant, we find an error threshold of $\sim16\%$, which is much higher than the threshold for spin-based topological memories like the Surface code or the Color code. In addition, we derive new lattice gauge theories to account for measurement errors. These results, together with the inherent error suppression provided by the superconducting gap in physical realizations of the MSC, makes this a strong candidate for a robust topological quantum memory.

\end{abstract}


\maketitle


\section{Introduction}
\label{sec:I}

Topological quantum memories (TQMs) \cite{Dennis_et_al02,Terhal_2015} provide a robust way to store quantum information in global (topological) properties of interacting many-body systems, where a larger system size implies a longer code distance. The non-local character of these phases of matter makes topological memories intrinsically robust against local perturbations and noise, which is key in the future development of quantum technologies. 
Topologically ordered systems in two spatial dimensions are characterized by a set of remarkable properties, namely, locally indistinguishable, degenerate ground states (where the quantum information is stored) and anyon excitations with fractional statistics.  For Abelian anyons, it is possible to perform logical gates by means of externally tailored operations depending on the physical system \cite{Kitaev97,Bombin07}. When anyons are non-Abelian, braiding of such anyons provides a toolbox of quantum gates to implement topologically-protected quantum computation \cite{Nayak_08}.


The vast majority of TQMs in the literature are spin-based models where errors have purely bosonic character, such as spin flips and phase errors. The paradigmatic example would be the \emph{Surface Codes} \cite{Kitaev_03,Bravyi_Kitaev_98}, constructed out of plaquette and star operators involving 4-body spin interactions on a square lattice. In order to allow for the transversal implementation of the whole Clifford group of quantum logic gates \cite{NC} a different class of bosonic topological codes was proposed, the \emph{Color Codes} \cite{colorcode1,colorcode2}. Remarkably, experimental realizations of small-scale instances of bosonic topological codes have been achieved in the lab \cite{colorqubit,surfacequbit,Gambeta15}.


More recently, TQMs based on Majorana fermions have been studied as a promising alternative to spin-based topological codes \cite{Bravyi_et_al10,Chiu_et_al10,MSC_1,MSC_2,Plugge_16,Landau_16,Vijay17,Hastings_17,Litinski_17,Roy_17,Constantin18_1,Constantin18_2,Litinski_18,Li16,Li18,Knapp18,Willie_et_al18}. Majorana devices benefit from the intrinsic robustness of unpaired Majorana particles and can be realized in a variety of platforms, such as the proximitized surface of a topological insulator \cite{bib:Fu2008,bib:Xu2015,bib:Sun2016}, semiconductor nanowires \cite{bib:Lutchyn2010,bib:Oreg2010,bib:Mourik2012,bib:Das2012,bib:Rokhinson2012,bib:Deng2013,bib:Deng2018,bib:Lutchyn2018}, and atomic arrays \cite{bib:Klinovaja2013,bib:Braunecker2013,bib:Pientka2013,bib:Nadj-Perge2013,bib:Nadj-Perge2014,bib:Ruby2015,bib:Pawlak2016}. Despite the intrinsic protection given by the superconducting gap, the fermionic character of these systems makes them susceptible to error channels that have no analog in spin-based memories, where spin flips or phase errors are the most natural sources of noise. Topological codes, constructed using interacting Majorana fermions, provide an additional layer of protection by means of redundancy. The \emph{Majorana Surface Code} (MSC) \cite{Bravyi_et_al10,MSC_1} is a particularly interesting stabilizer code in this class, with some remarkable properties over bosonic codes, such as single-shot readout of stabilizer measurements and lower overhead, since fewer physical qubits are required per encoded logical qubit  \cite{MSC_1}. Experimentally, they can be realized using charging energy-induced quantum phase slips in superconducting arrays with Majorana fermions. The two kinds of spontaneous errors that affect the code are quasiparticle poisoning (QP) and bilinear errors. QP is a purely fermionic error which changes the fermion parity associated with Majorana degrees of freedom \cite{Fu_09} and can cause topological qubits to decohere \cite{Chamon11,Loss_12}. This type of error is caused by an electron tunneling between the system and unknown states in the environment. In contrast, bilinear errors do not change the local fermion parity. 
Recent experiments \cite{Albrecht_17,Veen_18} have shown and measured how detrimental QP is in Majorana devices, with poisoning rates of the order of nanoseconds \cite{Veen_18}. This potential roadblock for Majorana-based technologies requires creative strategies in order to mitigate this experimental obstacle \cite{Karzig_17,Vijay17,Menard_19}. Hence, identifying these strategies is one of the most extraordinary challenges that solid-state quantum computers face, and it is precisely the main motivation for this article.

Traditionally, in order to protect stored quantum information from decoherence, quantum error-correcting codes have been proposed to actively detect and correct errors \cite{Kitaev97,Shor2,Steane,Gottesman_96,Preskill98,Terhal_2015,Vijay17,Li18,Knapp18}. For the conventional surface code, error detection occurs through local measurements \cite{Dennis_et_al02}, and the error threshold -- the maximum tolerable error rate, above which it is impossible to reliably encode quantum information -- can be mapped to the critical point at which the ordered phase of a dual statistical mechanical system is destroyed by the action of quenched disorder. 

In this paper, we demonstrate that a mapping to a classical statistical model can be used to extract the error-correcting properties of the MSC. We show that the error model for the MSC with quasi-particle poisoning maps to a random 3-body Ising model on a triangular lattice. The error threshold ($\sim11\%$) is the same as for both the Surface code \cite{Dennis_et_al02} and the Color code \cite{Katzgraber_et_al09}. In physical realizations of the MSC, quasi-particle poisoning is exponentially suppressed due to the presence of a superconducting gap and finite charging energy. In that case, when fermionic bilinear errors dominate, the MSC error model can be mapped to a collection of three random 2-body Ising models on triangular sublattices. The error threshold for this model is much higher ($\sim16\%$) than in the conventional surface code, which demonstrates that the MSC is more robust to bosonic errors (fermionic bilinears) than previously studied spin-based topological codes. In addition, accounting for measurement errors \cite{Ohno_04,Andrist11} yields lattice gauge theories that have not been previously studied. 

The paper is structured as follows. In Sec. \ref{sec:II}, we introduce the basics of the MSC. In Sec. \ref{sec:III}, we present the error models and elaborate on their physical motivation. In Sec. \ref{sec:IV}, we explain how the mapping to classical statistical models is performed and in Sec. \ref{sec:V}, we compute and compare the different error thresholds for the different operating regimes of the MSC. In Sec. \ref{sec:VI}, we derive lattice gauge theories that can be studied in order to understand error-thresholds in the presence of measurement errors.

\begin{figure}
\includegraphics[width=\columnwidth]{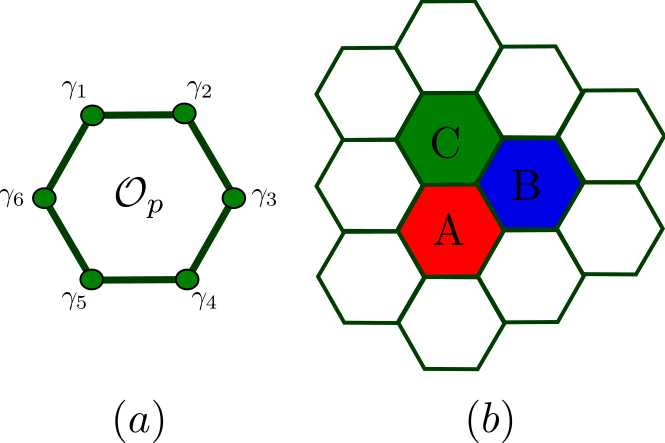}
\caption{(a) Honeycomb lattice with a single Majorana fermion per site. The plaquette operator $\mathcal{O}_{p}$ in the MSC is the product of the six Majorana fermions at the vertices. The colored plaquettes in (b) correspond to the three distinct bosonic excitations that may be generated by the action of a Majorana operator $\gamma_i$.}
\label{fig:Plaquette_1}
\end{figure}

\section{Majorana Surface Code}
\label{sec:II}
The Majorana Surface Code (MSC) \cite{MSC_1} is defined over 3-valent lattices with 3-colorable faces. The simplest instance consists of a honeycomb lattice with one Majorana fermion ($\gamma_i$) on each lattice site $i$. The Majorana operators satisfy the canonical anti-commutation relations $\{\gamma_{i},\gamma_{j}\} = 2\delta_{ij}$. The Hamiltonian is defined as the sum of operators acting on each hexagonal plaquette:
\begin{align}
H = -u\sum_{p}\mathcal{O}_{p} \hspace{.25in} \mathcal{O}_{p} \equiv i\prod_{n \in\rm{vertex} (p)}\gamma_{n}, 
\label{H1}
\end{align}
where $u>0$ and the operator $\mathcal{O}_{p}$ is the product of the six Majorana fermions on the vertices of plaquette $p$ as shown in Figure \ref{fig:Plaquette_1}a.  Since any two plaquettes on the honeycomb lattice share an even number of vertices, all of the plaquette operators (also called stabilizers) commute with each other and the Hamiltonian. The ground-state $\ket{\Psi_{0}}$ is defined by the condition 
\begin{equation}
\mathcal{O}_{p}\ket{\Psi_{0}} = \ket{\Psi_{0}},  \label{p}
\end{equation}
for all plaquettes $p$, i.e. the state that is left invariant by all stabilizers with eigenvalue $+1$. This ground state $\ket{\Psi_{0}}$ exhibits ${\cal Z}_2$ fermionic topological order. For an $N$-site honeycomb lattice, the $2^{N/2}$-dimensional Hilbert space of Majorana fermions 
is constrained by the system fermion parity:
\begin{align}
\Gamma \equiv i^{N/2}\prod_{n}\gamma_{n}. 
\label{gamma1}
\end{align}
The $N/2$ factor appears because one regular fermion is made out of two Majorana fermions. The conservation of the fermion parity $\Gamma$ is a unique property of Fermi systems, which separates two disconnected sector of excitations. Since the logic qubits are stored in the ground state subspace, excitations over the ground state manyfold account for errors in the quantum code. These errors can be divided into two groups, depending on whether they commute with the system's fermion parity in Eq. \eqref{gamma1} or not. The ground state constraints of the MSC are discussed in the Appendix.

\begin{figure}
\includegraphics[width=\columnwidth]{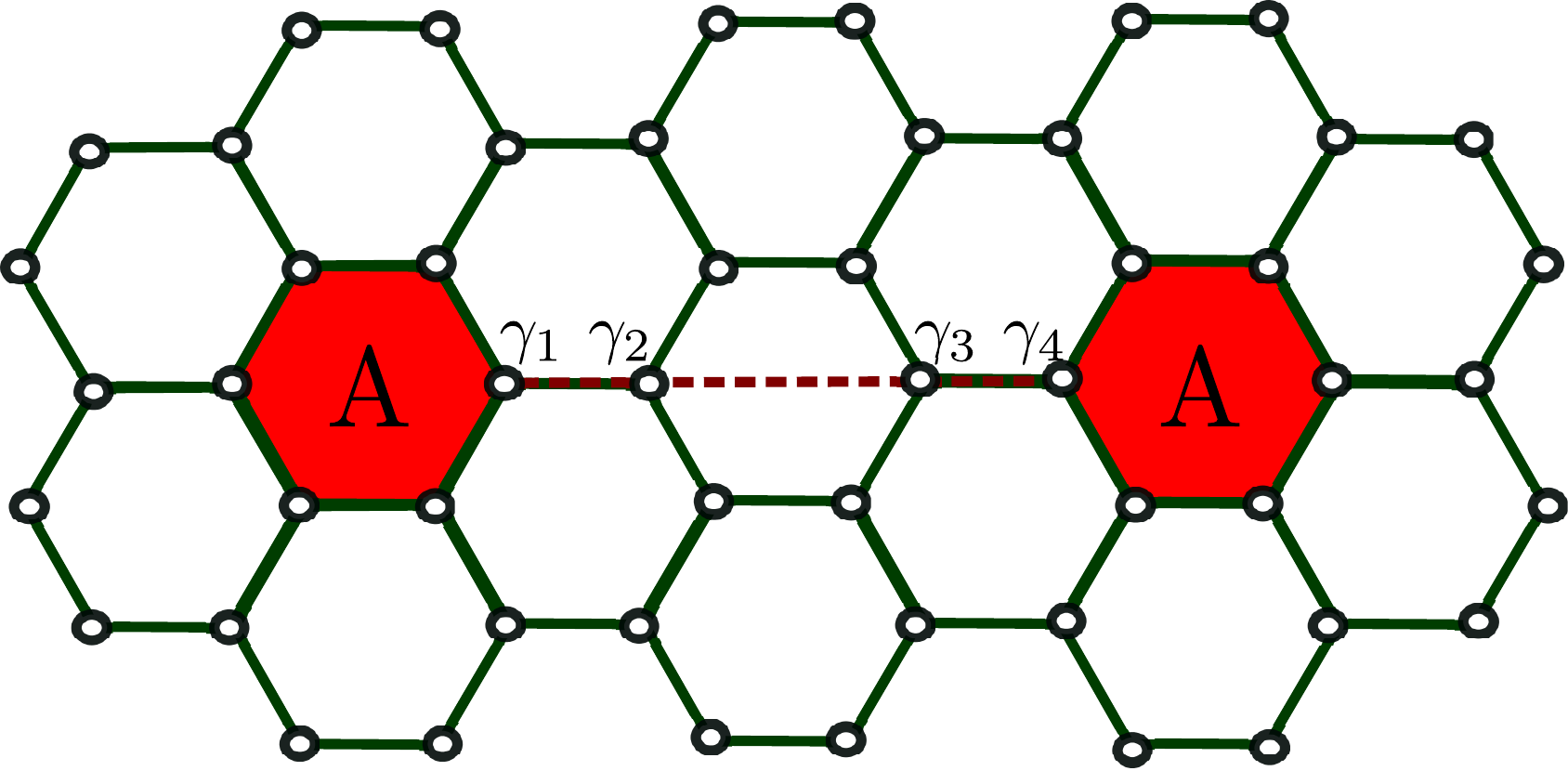}
\caption{An A-A plaquette-type excitation produced by the consecutive action of two bilinear errors ${\rm i}\gamma_1\gamma_2$ and ${\rm i}\gamma_3\gamma_4$.}\label{BiliniarAA}
\end{figure}

\section{Error Model}
\label{sec:III}

A quantum depolarising channel for a spin-based quantum memory is a completely positive trace-preserving map of a quantum state $\rho$ into a mixture of $\rho$ (the reduced density matrix of each single spin) and the maximally mixed state ${\mathcal I}$ (identity map) \cite{NC}. It is convenient to express this error channel in terms of the Pauli operators $\sigma^{x,y,z}$,
\begin{equation}
{\cal D}_{s}(\rho)=(1-p)\rho+\frac{p}{3}\sum_{r=x,y,z}\sigma^r\rho\sigma^r,
\end{equation}
where $\sigma_x$ models a spin flip error, $\sigma_z$ a phase flip error, and $\sigma_y$ a combination of both, occurring with probability $\frac{p}{3}$. The error $\sigma_y$ accounts for a correlated effect of spin-flip and phase errors \cite{Bombin_12}. This single-particle error channel neglects correlated error processes involving multiple spins, which may become relevant in multi-qubit gate implementations for quantum computation \cite{Raussendorf07,Wang11}. 

In the case of quantum memories constructed out of Majorana fermions, the physically motivated error channels are intrinsically different because of the fermionic nature of the system. Experimentally, two error sources are expected to play a major role: quasi-particle poisoning (QP) and bilinear errors.

QP errors are excitations over the ground state $\ket{\Psi_{0}}$ generated by operators of the form $\gamma_i$. They can be visualised as a violation of three neighboring plaquettes (ABC), as shown in Fig.~\ref{fig:Plaquette_1}b or the upper part of Fig.~\ref{mapping1}. They are purely fermionic errors, which change the fermion parity of the encoded qubit, and are caused by a single electron tunneling between the system and unknown states in the environment. In practise, they can be generated by low-temperature residual quasiparticles in the host superconductor tunneling into the two dimensional electron gas (2DEG) or topological insulator (TI) where the Majorana fermions are induced \cite{Loss_12,Albrecht_17,Karzig_17}.

Bilinear errors are excitations generated by operators of the form ${\rm i}\gamma_i\gamma_m$ acting on adjacent vertices. They conserve the total fermion parity $\Gamma$, and they can be visualised as two plaquette violations of the same type (A-A, B-B, C-C) joined by a string. In Fig.~\ref{BiliniarAA} we depict two A-plaquette violations. These errors can arise from wavefunction overlap of adjacent Majorana fermions or sequential QP events. 

\begin{figure}
\includegraphics[width=\columnwidth]{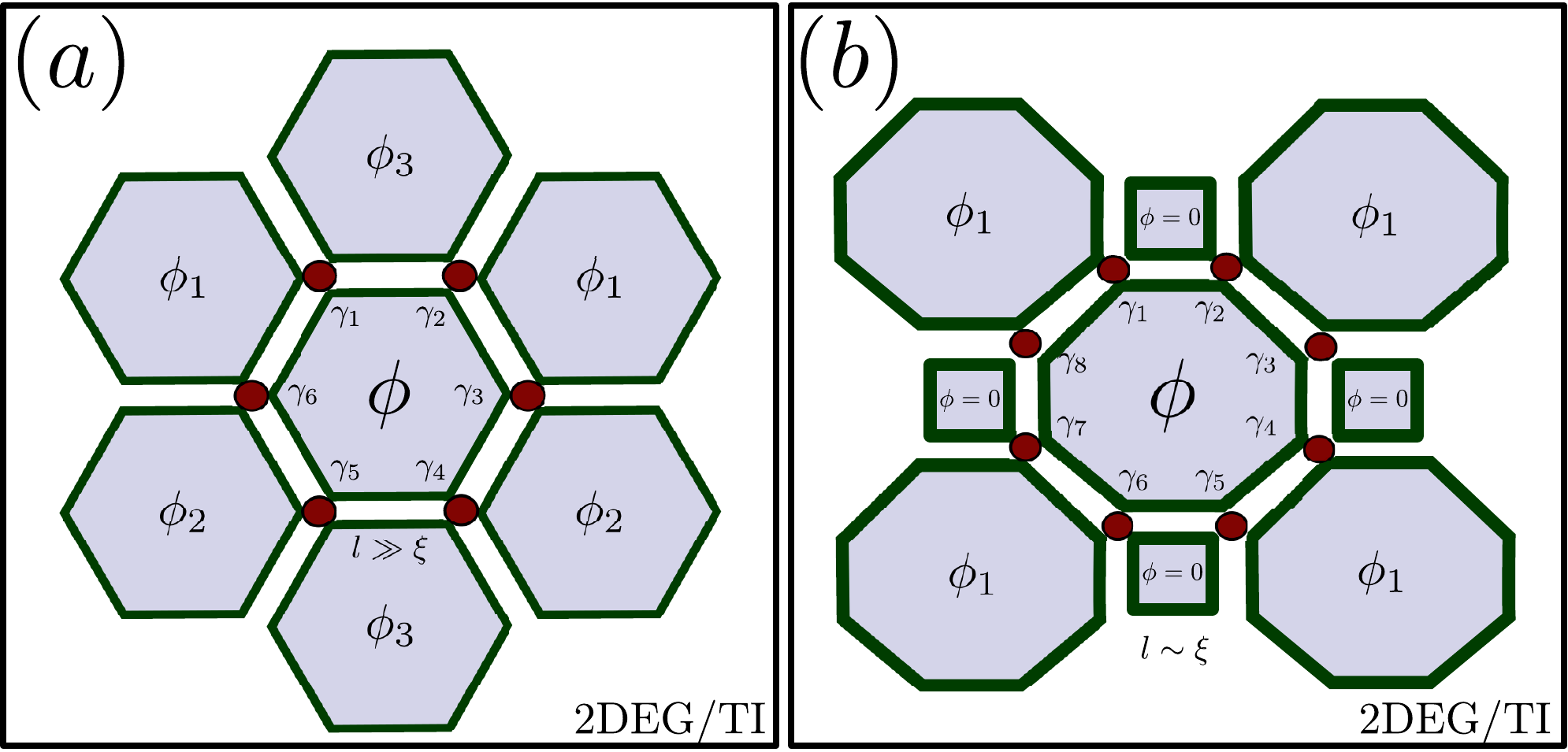}
\caption{The experimental platform consists of superconducting islands deposited on top of a TI surface or a 2DEG with strong spin-orbit coupling and a Zeeman field. (a) A phase-slip on the central hexagonal island generates a six-body interaction between the Majorana fermions $\gamma_i$. In the limit where the separation between different $\gamma_i$ is much larger than the coherence length $l\gg\xi$, QP would dominate over bilinear errors. (b) For a square/octagon lattice, the wavefunction overlap between each pair of $\gamma_i$ generates the eight-body plaquette terms in perturbation theory, provided that $l\sim\xi$. In this case, bilinear errors would dominate over QP events.}
\label{fig:2deg}
\end{figure}

Both types of errors can become the dominant source of noise depending on the particular experimental realization of the MSC in the lab. One realization \cite{MSC_1} involves Josephson-coupled mesoscopic topological superconducting islands, e.g. hexagonal s-wave superconducting islands imprinted on a TI surface or a 2DEG with strong spin-orbit coupling and subject to a Zeeman field, as shown in Fig. \ref{fig:2deg}a. The interacting plaquette terms ${\cal O}_p$ are generated through charging energy induced quantum phase-slips on the hexagonal islands. Since the separation between the Majorana fermions $l$ is much larger than the superconducting coherence length $l\gg\xi$, and the charging energy $E_c$ is small compared to the Josephson energy, QP events dominate over residual bilinear interactions. In square/octagon lattice proposals \cite{MSC_2}, a large $E_c$ suppresses QP and perturbatively generates the necessary eight-body plaquette operators from wavefunction overlap between Majorana fermions ($l\sim\xi$), as shown in Fig. \ref{fig:2deg}b. In such a situation, bilinear errors ${\rm i}\gamma_i\gamma_j$ would dominate over single-electron tunneling $\gamma_i$. These experimental realizations illustrate situations where either QP or bilinear errors can be the dominant error process.

Accounting for both types of errors, the completely positive trace-preserving map for the MSC casts the form 
\begin{eqnarray}
{\cal D}_{M}(\rho)&=&(1-p_{qp}-p_b)\rho + \\
&+& \frac{p_{qp}}{2}\gamma_i\rho\gamma_i + \frac{p_{qp}}{2}\gamma_j\rho\gamma_j + p_b\gamma_i\gamma_j\rho\gamma_j\gamma_i, \nonumber
\label{Dnoise}
\end{eqnarray}
where $\rho$ is the reduced density matrix for a pair of neighboring Majorana fermions ${\rm i}\gamma_i\gamma_j$. In the next section, we analyze the action of both fermionic and bosonic errors \cite{Vijay17} and we quest for the survival of quantum information. We show how this problem can be mapped to the phase transition determination of classical spin models.

\section{Mapping to classical spin models}
\label{sec:IV}

Computing the maximum error rate $p_c$ at which quantum error correction is still possible represents a cornerstone for both short-term and long-term quantum computing devices. In the case of topological quantum codes, this problem can be solved by finding the phase transition point of a classical spin system with quench disorder \cite{Dennis_et_al02}. The mathematical mapping establishing this surprising equivalence can be understood as follows.

\begin{figure}[t]
\centering
\includegraphics[width=\columnwidth]{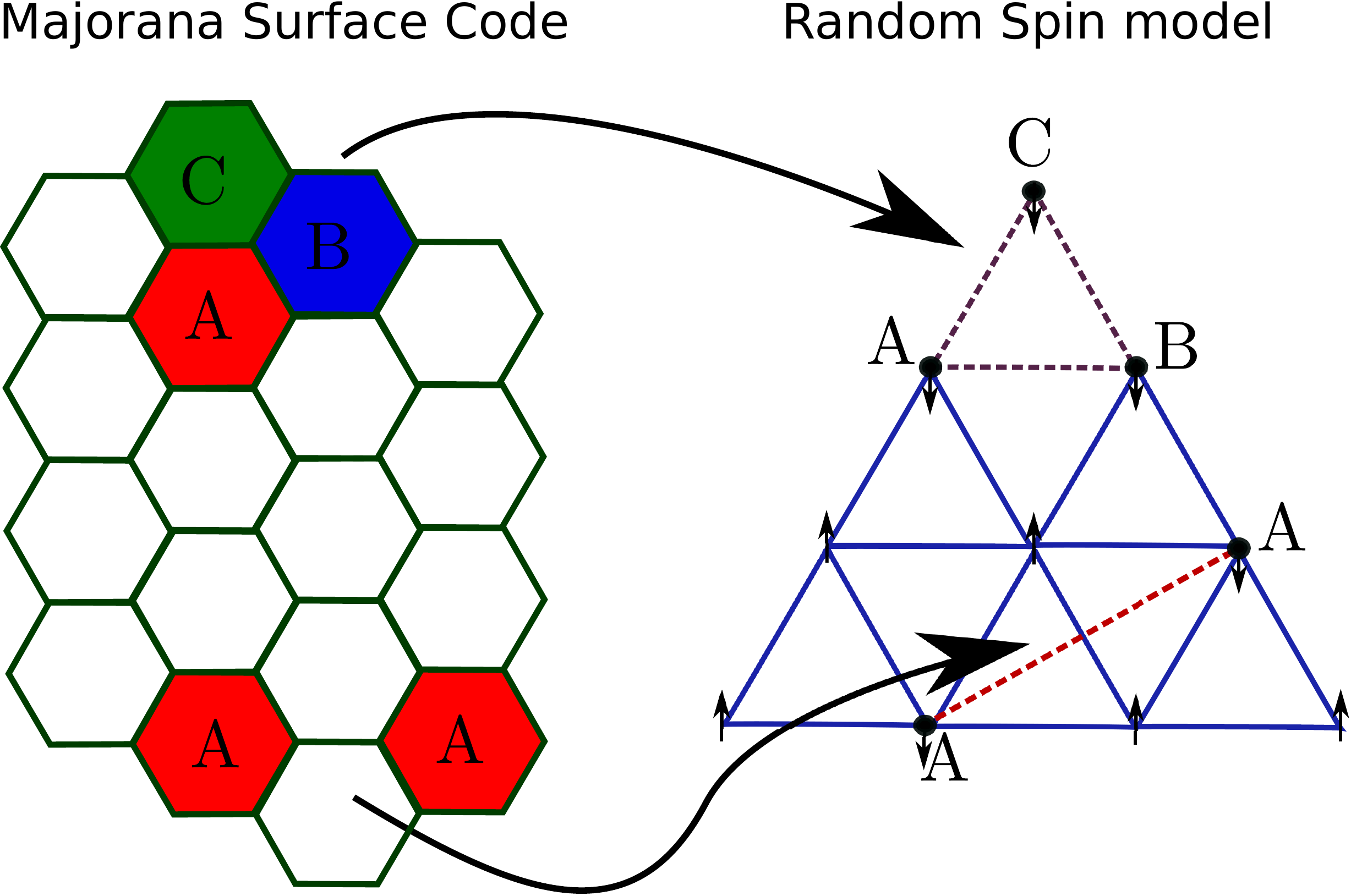}
\caption{Mapping from the Majorana Surface Code with errors to a classical random spin model on a triangular lattice, where interactions are represented by dashed lines. Quasi-particle poisoning errors (upper part) are mapped
to 3 body nearest-neighbor spin interactions. Bilinear bosonic errors (lower part of the figure) map to next-to-nearest neighbors spin interactions.}
\label{mapping1}
\end{figure}

Excitations (code errors) of Hamiltonian \eqref{H1} can be identified with violations of the plaquette ground state condition in Eq.~\eqref{p}, in the form of $\mathcal{O}_{r}\ket{\Psi} = -\ket{\Psi}$ for the affected plaquettes $r$. Therefore, each hexagonal plaquette can be labeled with a classical spin variable $s_{r}\in\{+1,-1\}$ at the vertices of the dual triangular lattice (see Fig. \ref{mapping1}) indicating the \emph{syndrome}, i.e. whether an error occurs on a plaquette. 

When these errors concatenate, they form an error chain $E$. From a practical quantum error correction point of view, this error chain will lead to plaquette violations only at the endpoints (see Fig. \ref{BiliniarAA}), which means that any other error $E'$ with the same endpoints would give the same error syndrome. All equivalent errors, with equal endpoints, belong to the same error class $\hat{E}$.

A successful error correction strategy can only take place if correcting for an error class $\hat{E}$ is independent of whether the physical error chain was $E$ or any other error chain $E'$ within the same error class. For topological stabilizer codes this is precisely the case. This apparently surprising fact can be explained because the difference $E'-E=C$ is a trivial homological cycle, i.e. a surface without a boundary formed by a product of stabilizer operators $\mathcal{O}_{r}$, which leaves the state of the system invariant. Thus, correcting for $E$ is equivalent to correcting for $E'$ if $\{E,E'\}\in\hat{E}$.
This condition is satisfied as long as the difference $C=E'-E$ never comprises a non-trivial homological cycle that would change the ground state of the system and consequently corrupt the stored quantum information. This requirement implies that error chains should not proliferate for quantum error correction to be feasible. 

By computing the relative probability of error $E'$ with respect to error $E$ occurring in the system, and sampling over all possible configurations of error chains, we obtain a classical partition function $Z=\sum_{\{s\}} {\rm exp}\big(-\beta H\big)$, where $\beta$ stands for the inverse temperature, and $H$ is the Hamiltonian of a classical spin model \cite{Dennis_et_al02}. The inverse temperature $\beta$ is not the physical temperature of the quantum memory, but an ancillary parameter of the associated classical model that is used for statistical sampling. The eigenvalues of each stabilizer $O_r$ of the MSC can be mapped to Ising spins with interactions dictated by how each stabilizer is affected by errors on adjacent Majoranas. The associated classical Hamiltonian casts the general form \cite{Katzgraber_et_al09}:
\begin{equation}
\label{general_ham}
{\cal H}=-J\sum_{l}\sum_{w}\tau_l^w\prod_is_i^{g_{il}^w},
\end{equation}
where $l$ runs over all qubit sites, $i$ runs over all stabilizers, $s_{i}$ is a classical spin variable representing the eigenvalue of the associated stabilizer $\mathcal{O}_{i}$, $w$ indicates the type of error (quasi-particle poisoning, bilinear, etc), $g_{il}^w\in\{0,1\}$ determines whether the stabilizer $\mathcal{O}_{i}$ is affected by an error of type $w$ acting on qubit $l$,
\begin{equation}
\label{weights}
\tau_l^w=
  \begin{cases}
    +1       & \quad \text{with~probability}~1-p\\
    -1  & \quad \text{with~probability}~p
  \end{cases}
\end{equation}
where $p$ is the probability of an error of type $w$ to occur, and the coupling constant $J$ is fixed by the so-called Nishimori condition \cite{Nishimori80,Nishimori_81},
\begin{equation}
-2\beta J=\text{ln}\Big(\frac{p}{1-p}\Big).
\label{Nishi}
\end{equation}

Therefore, we assign an antiferromagnetic interaction to the classical spins $s_r$ (representing the eigenvalues $+1,-1$ of stabilizers ${\cal O}_r$) affected by the sampled error class, a ferromagnetic interaction to spins not affected by that error class, and we sample over all error classes. Thus, the ferromagnetically ordered phase in the classical spin model maps to the phase where errors are bounded by domain walls and do not proliferate. This is the situation where we can correct errors successfully. 

In the following section we compute the optimal error thresholds for all possible error channels in the MSC and discuss how they compare to the thresholds for the more studied bosonic topological codes.

\section{Fermionic and bosonic error correction}
\label{sec:V}

Code errors in the MSC can be divided into fermionic or bosonic, depending on whether they conserve the total fermion parity of the system. We analyze both error channels and argue how the MSC outperforms standard bosonic topological codes, like the Surface Code \cite{Kitaev_03,Bravyi_Kitaev_98} and the Color code \cite{colorcode1,colorcode2}.

\subsection{Quasi-particle poisoning}

QP is a purely fermionic error, which does not commute with the fermion parity operator of the system (see Eq.\eqref{gamma1}). It is created by one-body Majorana operators $\gamma_i$ acting on single vertices $i$, and affects the three surrounding plaquette operators labeled by classical spin variables $s_i$.
Applying these constraints to Eq.\eqref{general_ham}, the corresponding classical spin Hamiltonian is a two-dimensional 3-random-bond Ising model on a triangular lattice, 
\begin{equation}
{\cal H}=-J \sum_{ijk\in{\Delta}}\tau_{ijk}s_is_js_k,
\label{Z1}
\end{equation}
with $\tau^p_{ijk}$ randomly chosen between $\{+1,-1\}$, such that the probability to pick $\tau_{ijk}=-1$ is given by the quasi-particle poisoning error probability $p$. The global coupling constant $J$ is fixed according to the Nishimori condition \cite{Nishimori80} as given in Eq.\eqref{Nishi}.

In order to determine the error threshold $p_c$ for QP, the critical temperature $T_c(p)$ as a function of $p$ is computed for the classical spin model in Eq.~\eqref{Z1}. Secondly, the crossing point of this curve $T_c(p)$ with the Nishimori line given in Eq.~\eqref{Nishi}, yields the error threshold $p_c$. 

Hamiltonian \eqref{Z1} is equivalent to the classical spin model given by the Color Code on a hexagonal lattice with only qubit-flip errors \cite{Bombin_08,Katzgraber_et_al09}. The error threshold $p_c=0.109(2)$ was obtained using Monte Carlo simulations \cite{Katzgraber_et_al09} and also coincides with the error threshold for the Surface Code with qubit-flip errors \cite{Dennis_et_al02}. This result shows that Majorana-based topological codes are robust to purely fermionic errors below large thresholds of $\sim11\%$.

\begin{figure}[t]
\centering
\includegraphics[width=\columnwidth]{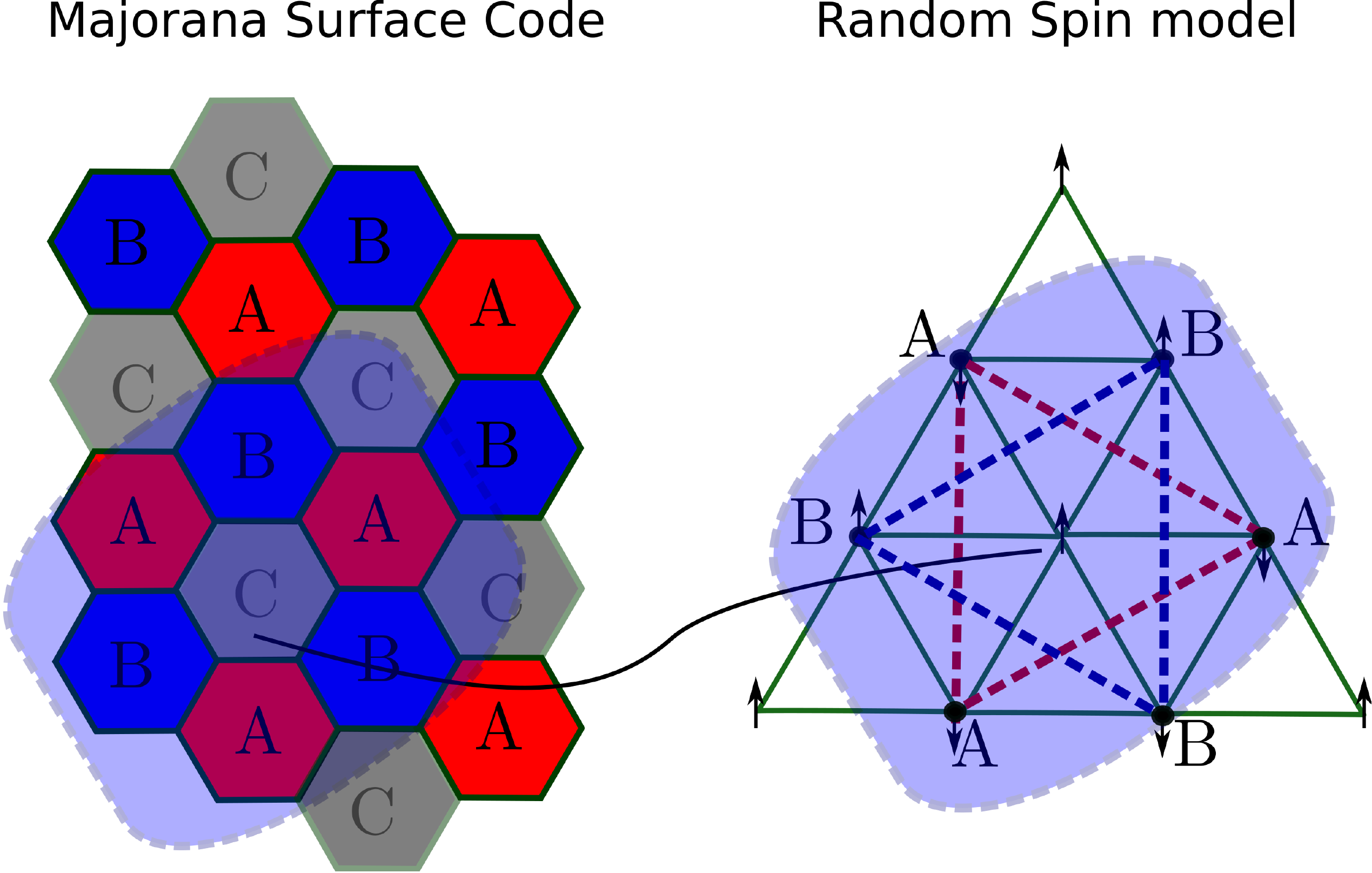}
\caption{Mapping of $A-A$ and $B-B$ bilinear errors to a classical spin model. Increasing the charging energy of the $C-$plaquettes with respect to $A$ and $B$ types, we can prevent $C-C$ bilinear errors. Pairs of plaquette excitations of type $A(B)$ can be understood as nearest neighbor spin interactions over independent triangular sublattices.}
\label{bilinear}
\end{figure}

\subsection{Bilinear errors}

Bilinear fermionic errors (bosonic) do not change the fermion parity, and are created by the action of operators ${\rm i}\gamma_i\gamma_j$ acting on adjacent sites. They generate plaquette violations of the same type, in pairs ($A-A$, $B-B$ and $C-C$), which do not get mixed with each other as shown in Fig.~\ref{BiliniarAA}. The associated classical spin model corresponds to 3 independent nearest-neighbor 2-random-bond Ising models over a triangular lattice (see Fig.~\ref{bilinear}). The classical spin Hamiltonian is obtained by applying the previous constraints to Eq.~\eqref{general_ham}, which yields
\begin{equation}
{\cal H}= -K \sum_{<ij>}\tau_{ij}s_is_j,
\label{Hb}
\end{equation}
with $\tau_{ij}$ randomly chosen between $\{+1,-1\}$, such that the probability to pick $\tau_{ij}=-1$ is given by the bilinear error probability $q$ of the MSC. The coupling constant $K$ is fixed according to the Nishimori condition in Eq.~\eqref{Nishi}. Due to homogeneity of the space, the error probability $q$ is the same for the three sublattices, and since they do not mix with each other, the error threshold $q_c$ will be the same for all of them.

This classical statistical model given in Eq.~\eqref{Hb} has been previously studied in the community of spin glasses, where $1-q_c$ was reported to be 0.8355(5) \cite{Queiroz_06}, which makes the error threshold $q_c=0.1645(5)$. This error threshold is almost $50\%$ higher than for spin-based topological codes with either spin-flips or phase errors. Such high tolerance to errors in the MSC ($\sim16\%$) makes it a strong candidate for a robust topological quantum memory.

\subsection{Combined Errors}

By taking into account both quasi-particle poisoning and bilinear errors, we obtain a novel classical spin model, which up to our knowledge has been never considered before:
\begin{equation}
{\cal H}=- J \sum_{ijk\in{\Delta}}\tau^p_{ijk}s_is_js_k -K \sum_{\langle\langle ij\rangle\rangle}\tau^q_{ij}s_is_j,
\label{H3}
\end{equation}
where the bilinear interaction $s_is_j$ acts only over next-to-nearest neighbor classical spins corresponding to the same type of plaquette, as shown in Fig.~\ref{mapping1}. The coupling constants $\tau^{p}_{ijk}$ and $\tau^q_{ij}$ are randomly chosen between $\{+1,-1\}$, such that the probability to pick $\tau^p_{ijk}=-1$ is given by $p$ (the quasi-particle poisoning error probability of the MSC) and the probability to pick $\tau^q_{ij}=-1$ is given by $q$ (where $q$ is the bilinear error probability of the MSC).
The coupling constants $J$ and $K$ are now fixed by two Nishimori conditions
\begin{equation}
{\rm e}^{-2\beta J} = \frac{p}{1-p}~~~,~~~{\rm e}^{-2\beta K} = \frac{q}{1-q}.
\label{N3}
\end{equation}
The choice of $q$ and $p$ should be based on the particular experimental proposal to realised the MSC \cite{MSC_2}. We have considered QP and bilinear errors independently since they arise from physically distinct error processes.

\section{Measurement Errors}
\label{sec:VI}

An implicit assumption of the error correcting model used thus far is that in order to detect an error chain $E$, the eigenvalues of the plaquette operators ${\cal O}_p$ have to be correctly measured (perfect syndrome measurement). For realistic applications, it is natural to also consider faulty Majorana stabilizer measurements. With that purpose in mind, we define the concept of \emph{error history}, which comprises the set of both faulty Majoranas and stabilizer error measurements. These error histories can be classified into equivalence classes $\overline{E}$ (those that differ by only local equivalences), in such a way that correcting an error chain $E\in\overline{E}$ leads to successful recovery, regardless of whether a different error $E'\in\overline{E}$ occurred instead. This statement holds as long as both errors belong to the same error class $\overline{E}$. Measurement errors are incorporated into the error description by considering vertically stacked lattices: one layer with hexagonal colored plaquettes (the MSC) followed by the dual lattice with triangular plaquettes (corresponding to the measurement results of each stabilizer), and so on. 

A specific error history $E$ can be represented by a set of variables $\tau_l\in\{+1,-1\}$, each indicating whether a Majorana fermion at site $l$ or a measurement of a nearby stabilizer was faulty. In addition, we can enumerate all histories within the same error class by attaching binary variables $s_i\in\{+1,-1\}$ to each equivalence. In order to sample from these error equivalences, we construct a classical Hamiltonian whose partition function has Boltzmann weights proportional to the probability of, given an error chain $E$, finding another error $E'$ within the same equivalence class. Note that error histories in the same class can only differ by elementary equivalences, i.e., histories which have no boundary and are homologically trivial. 

This construction is just a generalisation of our previous discussion on error chains without faulty measurements. In the previous case, local equivalences were given by stabilizer operators (since acting with a stabilizer operator on the code leaves any quantum state invariant). Once we include measurements, there are two possible local equivalences: 

\textbf{Spatial-like}: Acting on the code with a plaquette operator $O_p$ does not change the state of the system. Therefore, two errors $E$ and $E'$ differing only by a set of plaquette operators are equivalent. Indeed we can associate a binary variable $s^{S}_i\in\{+1,-1\}$ to each plaquette equivalence. 

\textbf{Time-like}: these equivalences arise within the time-propagation of errors, and they involve both faulty Majorana fermions and stabilizer measurements.  
The first equivalence is defined by the action of a Majorana operator $\gamma_j$, which gets undetected by the 3 surrounding stabilizer operators, and another Majorana operator $\gamma_j$ is applied on the same site. We associate a binary variable $s^{T}_i\in\{+1,-1\}$ to each of these 3D trigonal bipyramidal volumes. 
The second equivalence involves a bilinear operator acting on two adjacent Majorana $\gamma_i\gamma_j$, which is undetected by 2 stabilizer operators, and a consecutive bilinear operators is applied on the same Majorana pair. We associate a binary variable $s^{T}_i\in\{+1,-1\}$ to each of these new plaquettes.

Now we analyze the classical error model associated to a combination of spontaneous and measurement errors, defined over a 3D lattice of stacked hexagonal and triangular 2D sublattices. 

\subsection{QP $\&$ Measurement Errors}
If the system is subject to QP and faulty measurements, the associated classical Hamiltonian is
\begin{equation}
{\cal H}=- J \sum_{j\in{\cal Q}}\tau^{q}_{j}(s^{S})^{\otimes 3}(s^{T})^{\otimes 2} -K \sum_{j\in{\cal M}}\tau^{m}_{j}(s^{T})^{\otimes 6}, 
\label{H4}
\end{equation} 
where ${\cal Q}$ iterates over all Majorana sites, and ${\cal M}$ iterates over all measurements. The first term in the Hamiltonian describes the equivalence classes affected by a single faulty Majorana fermion: 3 plaquette stabilizers (spatial-like equivalences) surrounding the faulty Majorana, and the upper or lower vertex of a trigonal bipyramidal time-like plaquette. The second term in Eq.~\eqref{H4} accounts for the equivalence classes affected by a faulty measurement: six trigonal bipyramidal time-like plaquettes altered by the faulty stabilizer ${\cal O}_p$. The $S$ spins belong to each triangular lattice layer, and the $T$ spins to the hexagonal lattice layers of the overall 3D lattice structure.
The coupling constants $\tau^{p}_{j}$ and $\tau^m_{j}$ are randomly chosen between $\{+1,-1\}$, such that the probability to pick $\tau^p_{j}=-1$ is given by $p$ (the QP error probability) and the probability to pick $\tau^m_{j}=-1$ is given by $m$ (the measurement error probability). The coupling constants $J$ and $K$ are fixed by their respective Nishimori conditions (Eq.~\eqref{Nishi}).

The resulting statistical model is a three-dimensional Ising lattice gauge theory with disorder, the same as the one obtained for the Color Code with qubit flips and measurement errors \cite{Andrist11}. Due to gauge symmetry, there is no local order parameter distinguishing the low temperature Higgs phase and the highly disordered confined phase. However, using the average value of a Wilson loop operator (non-local order parameter), and assuming equal error probabilities $p=m$, it is possible to retrieve the error threshold $p_c=0.048$, which is higher than the one for the Surface code $p_c=0.033$. This relatively higher error-threshold with respect to the Surface code is consistent for different values of $p$ and $m$ \cite{Andrist16}.

\subsection{Bilinear $\&$ Measurement Errors} 

If the system is subject to bilinear errors and faulty measurements, it maps to an associated classical Hamiltonian for each type of plaquette operator $A$, $B$ or $C$, 
\begin{equation}
{\cal H}=- J \sum_{j\in{\cal Q}}\tau^{b}_{j}(s^{S})^{\otimes 2}(s^{T})^{\otimes 2} -K \sum_{j\in{\cal M}}\tau^{m}_{j}(s^{T})^{\otimes 6}, 
\label{H5}
\end{equation} 
where ${\cal Q}$ iterates over all possible Majorana pairs connecting plaquettes of type $A,B$ or $C$; and ${\cal M}$ iterates over all $A,B$ or $C$-type stabilizer measurements. The first term in the Hamiltonian describes the equivalence classes associated to a faulty Majorana bilinear, i.e., the 2 plaquette stabilizers surrounding the bilinear and 2 time-like equivalences. A faulty Majorana bilinear can be the upper or lower edge of a time-like plaquette. The second term in Eq.~\eqref{H4} accounts for the equivalence classes involving a syndrome measurement error. Each plaquette measurement is involved in 6 time-like equivalences. The $S$ spins belong to each triangular lattice layer, and the $T$ spins to the hexagonal lattice layers of the overall 3D lattice structure.
The coupling constants $\tau^{b}_{j}$ and $\tau^m_{j}$ are randomly chosen between $\{+1,-1\}$, such that the probability to pick $\tau^b_{j}=-1$ is given by $b$ (bilinear error probability) and the probability to pick $\tau^m_{j}=-1$ is given by $m$ (the measurement error probability). The coupling constants $J$ and $K$ are fixed by their respective Nishimori conditions (Eq.~\eqref{Nishi}).
The resulting statistical model is a three-dimensional Ising lattice gauge theory with disorder, that up to our knowledge has never been studied before. The critical error threshold will be higher than the previous model for QP, since it involves the same lattice structure, but the degree of bond-interaction is lower, which usually implies less frustration and a higher critical temperature. Moreover, in the case of small $m$, we should recover the error threshold for bilinear errors of $\sim16\%$.

\section{Conclusions}
\label{sec:VII}

In this work, we have studied the Majorana Surface Code as a platform to perform fermionic and bosonic quantum error correction. After motivating the experimental relevance of these error sources, we have shown how purely fermionic errors, like QP, can be corrected as long as the error rate is below a threshold of $\sim11\%$. Moreover, we have demonstrated that bosonic errors in the MSC have a much higher error threshold (of $\sim16\%$) than spin-based TQMs like the Surface Code or the Color Code. Finally, we have derived new lattice gauge theories to account for measurement errors. 
The MSC is constructed placing Majorana fermions at fixed positions, whereas braiding proposals for topological quantum computation involve moving Majoranas around in complicated ways. Both computational models can implement the whole Clifford group \cite{colorcode2,Nayak_08,MSC_1} in a topologically protected way, which is not enough to perform universal quantum computation \cite{NC}. Extensions of spin color codes to three spatial dimensions (3D) were shown to allow for the transversal implementation of $T-$gates, needed for universal quantum computation \cite{Bombin_08}. Similar extensions of the MSC to 3D will be explored in a forthcoming publication.
\vspace{0.5cm}

\textcolor{white}{artificial space}

\begin{acknowledgements}

We thank M.A. Martin-Delgado for helpful discussions and a critical reading of the manuscript. This work was supported by the DOE Office of Basic
Energy Sciences, Division of Materials Sciences and Engineering
under Awards DE-SC0010526 and DE-SC0019275. OV thanks Fundacion Ramon Areces and RCC Harvard. SV is supported
by the Harvard Society of Fellows. LF is partly supported
by the David and Lucile Packard Foundation.

\end{acknowledgements}

\appendix
\setcounter{equation}{0}
\renewcommand*{\theequation}{A\arabic{equation}}
\section{Constraints for the MSC code}
\label{App_A}

We place the system defined in Eq.\eqref{H1} on a torus geometry by imposing periodic boundary conditions in both $x$ and $y$ directions.
For convenience, we choose the honeycomb lattice unite cell consisting of three plaquettes labeled $A, B$ and $C$, as shown in Figure \ref{fig:Plaquette_1}b. On the torus, the product of plaquette operators on each of the $A$, $B$ and $C$-type plaquettes is equal to the total fermion parity: 
\begin{align}
\Gamma = \prod_{p\in A}\mathcal{O}_{p} = \prod_{p\in B}\mathcal{O}_{p} = \prod_{p\in C}\mathcal{O}_{p}.
\label{gamma2}
\end{align}
The operators $\mathcal{O}_{p}$ on any one type of plaquette fix one-third of the plaquette eigenvalues through Eq.\eqref{p}, and 
impose $2^{N/6 - 1}$ constraints on the Hilbert space. The number of unconstrained degrees of freedom is therefore given by:
\begin{align}
D = 2^{\frac{N}{2} - 1}/\left( 2^{\frac{N}{6} - 1}\right)^{3} = 4, 
\end{align}
which yields a four-fold ground state degeneracy for the MSC on the torus \cite{MSC_1}.


\end{document}